# Experiment-based deep learning approach for power allocation with a programmable metasurface

Jingxin Zhang,[1,†] Jiawei Xi,[1,†] Peixing Li,[2] Ray C. C. Cheung,[2] Alex M. H. Wong,[2,3,*] and Jensen Li[1,*]

[1] Department of Physics, The Hong Kong University of Science and Technology, Clear Water Bay, Kowloon, Hong Kong, China

[2] Department of Electrical Engineering, City University of Hong Kong, China

[3] State Key Laboratory of Terahertz and Millimeter Waves, City University of Hong Kong, China

* alex.mh.wong@cityu.edu.hk, jensenli@ust.hk

† These authors contributed equally to this work.

## Abstract

Deep learning, as a highly efficient method for metasurface inverse design, commonly use simulation data to train deep neural networks (DNNs) that can map desired functionalities to proper metasurface designs. However, the assumptions and simplifications made in the simulation model may not reflect the actual behavior of a complex system, leading to suboptimal performance of the DNNs in practical scenarios. To address this issue, we propose an experiment-based deep learning approach for metasurface inverse design and demonstrate its effectiveness for power allocation in complex environments with obstacles. Enabled by the tunability of a programmable metasurface, large sets of experimental data in various configurations can be collected for DNN training. The DNN trained by experimental data can inherently incorporate complex factors and can adapt to changed environments through its on-site data-collecting and fast-retraining capability. The proposed experiment-based DNN holds the potential for intelligent and energy-efficient wireless communication in complex indoor environments.

# 1. Introduction

Metasurfaces, consisting of subwavelength artificial array structures on an ultrathin surface, possess a remarkable ability to fully control the properties of electromagnetic (EM) waves, including their amplitude, phase, polarization, and wavefront structure. This control leads to the generation of exotic electromagnetic responses, such as a negative refractive index [1, 2], perfect absorption [3], superlensing [4], and invisibility cloaking [5, 6]. However, these passive metasurfaces are usually inherently limited to specific functions once their fabrication is finalized and thus cannot meet the requirement of dynamic control of EM waves. Recently, significant efforts have been devoted to developing active or reconfigurable metasurfaces [7-13], whose properties can be varied through external tuning. Particularly, programmable metasurfaces [14] as a cost-effective implementation of reconfigurable intelligent surfaces (RISs) [15-17] have emerged by digitally controlling the metasurface with the field-programmable gate array (FPGA), allowing for the manipulation of EM waves in both space and time. Programmable metasurfaces have demonstrated considerable potential for various applications, including scanning [18, 19], spatial frequency multiplexing [20-22], nonreciprocal reflection [23], holographic imaging [24-26], and orbital angular momentum generation [27, 28].

With growing interest in exploring new phenomena and more complex applications with metasurfaces, an efficient and accurate metasurface inverse design method has become crucial. Conventional inverse design procedures are usually guided by optimization algorithms [29-32], which are usually time-consuming iterative searching steps in a case-by-case manner. However, iterative methods cannot be implemented for real-time applications that require fast switching between different functionalities. Recently, deep learning approaches have been more efficient methods for metasurface inverse designs [33-38]. Once the deep neural network (DNN) is well-trained by vast amounts of data, the network can immediately find proper metasurface designs for different targets without going through iterations again. Particularly, the DNN-assisted metasurface inverse design has been applied in beamforming and power allocation [39-43] to control the power delivered toward target users at different locations for wireless communication systems. However, most of these works rely on training the DNN with simulation data, which may oversimplify the modeling parameters in complex scenarios, thus limiting its use in realistic situations [44, 45]. Additionally, changes in environments can significantly compromise the performance of a trained DNN or render it ineffective.

In this work, we propose an experiment-based deep learning approach for metasurface inverse design to achieve power allocation in complex environments with obstacles. Without the need for sophisticated modeling and time-consuming simulations, we train the DNN directly with experimental data, which can be measured in various configurations of a programmable metasurface enabled by its programmability. Experimentally collected data can inherently incorporate complex factors in realistic situations that are difficult to be included in the simulation model. Our results demonstrate that the experiment-based DNN is effective in controlling the power transmitted toward multiple receivers and can adapt to changed environments through its on-site data-collecting and fast-retraining capability. The proposed experiment-based deep learning scheme offers a promising direction for leveraging real-world data to achieve accurate and efficient metasurface inverse designs for boosting or damping Wi-Fi and 5G signals in complex indoor environments.

## 2. Methods
### 2.1 DNN for power allocation with programmable metasurface

We aim to control the power transmitted to specific receivers in a complex environment, generally with an obstacle, using a programmable metasurface together with an experiment-based deep learning approach, as shown in Fig. 1. A reflective programmable metasurface featuring tunable reflection phase profiles in the microwave regime is illuminated with a monochromatic excitation signal from a feed horn. The metasurface comprises 20 columns of unit cells, and the reflection phase $\{\varphi_i\}$ ($i = 1,2,...,20$) for each column can be independently controlled. After the reflected wave is scattered by an obstacle (a metal frame in this case), the scattered field intensities ($I_{m1}$, $I_{m2}$, and $I_{m3}$) are measured by three open-end waveguide probes in specific locations. Our deep neural network (DNN) consists of a forward scattering engine (FSE) and an inverse-design engine (IDE), as shown in Fig. 1. We first train the FSE in turning a set of reflection phases $\{\varphi_i\}$ into the predicted scattered fields $\{I'_j\}$ ($j = 1,2,3$). During the training, a large number of randomly generated configurations of $\{\varphi_i\}$ and the corresponding intensities of the experimentally measured scattered field $\{I_{mj}\}$ by the 3 probes are used as training data. The mean squared error (MSE) between $\{I'_j\}$ and $\{I_{mj}\}$ is used as the loss function to optimize the FSE during the training process. Specifically, the FSE is a supervised network with 40-100-100-3 fully connected layers. We opted to split the cyclic reflection phase $\{\varphi_i\}$ into $\{\cos\varphi_i, \sin\varphi_i\}$, resulting in 40 input variables for the 20 columns of reflection phases, which can improve training performance (see Sec. 2 in the Supplementary Materials). The FSE has two hidden layers both with 100 neurons, using the exponential linear unit (ELU) activation function, and has 3 output variables for the predicted intensities $\{I'_j\}$. After training, the FSE acts similarly as a surrogate solver in replacing full-wave simulations, except now replacing the real physical scattering process.

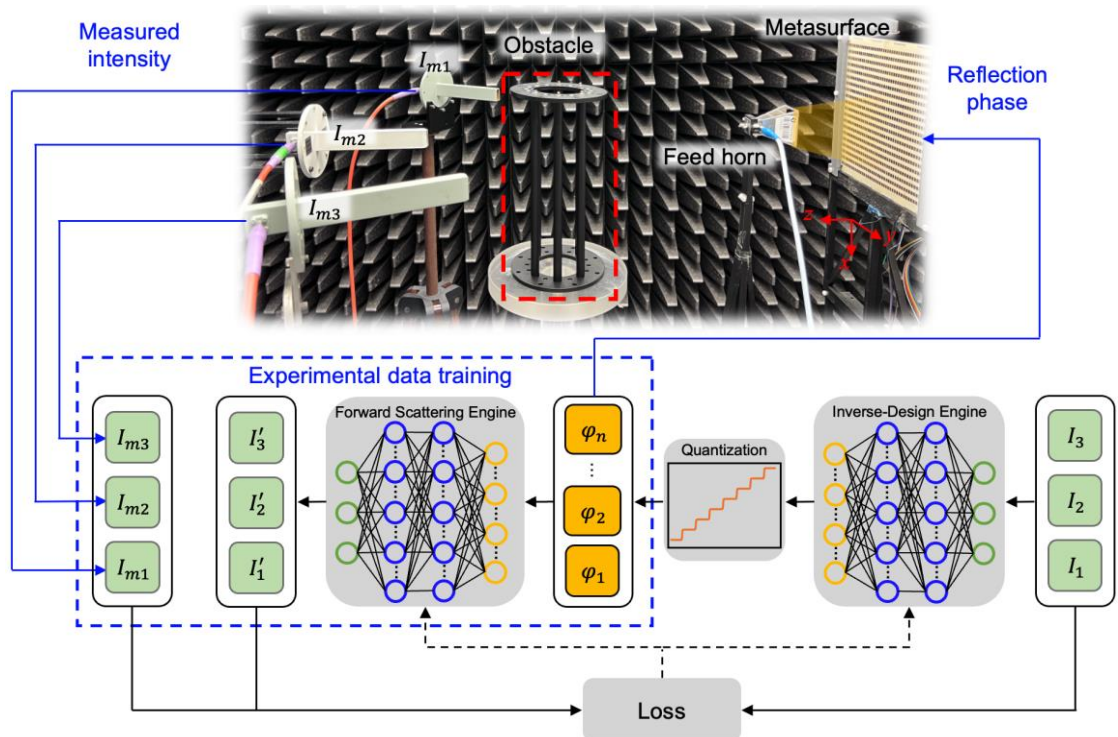

Fig. 1 Schematic of the experiment-based DNN for power allocation with a programmable metasurface. A reflective metasurface with tunable reflection phase profiles $\{\varphi_i\}$ is excited by a monochromatic source at 11GHz from a feed horn. Three fixed probes in specific locations are employed to measure the corresponding field intensities $\{I_{mj}\}$ after scattering by a metal frame obstacle. By collecting large sets of $\{I_{mj}\}$ as the experimental training data, a forward scattering engine (FSE) is pre-trained to convert reflection phase profiles $\{\varphi_i\}$ to predicted intensities $\{I'_j\}$. Additionally, an inverse-design engine (IDE) is developed to transform input target intensities $\{I_j\}$ to the output required reflection phase profiles $\{\varphi_i\}$. The integrated DNN, which combines the IDE with the pre-trained FSE, is trained to co-ordinate the metasurface inverse design to manipulate the scattered fields on demand.

Next, the IDE is constructed with reverse topology of 3-50-50-20 fully connected layers. The input target intensities $\{I_j\}$ are inversely transformed to the desired reflection phase profile $\{\varphi_i\}$. During the training the IDE, the MSE between $\{I_j\}$ and $\{I'_j\}$ (IDE combined with the pre-trained FSE) is used as the loss function and no experimental data is needed in this stage. Finally, for any target set of $\{I_j\}$, the output of the IDE, $\{\varphi_i\}$, can now be used as the input of the real metasurface to test whether the experimentally obtained $\{I_{mj}\}$ is similar to $\{I_j\}$. We note that due to the inverse design nature of the problem, there may be multiple phase profiles $\{\varphi_i\}$ that can achieve the same set of target intensities $\{I_j\}$. The integration of the IDE and pre-trained FSE as an integrated DNN (an autoencoder setting for results instead of design parameters) can help mitigate the non-uniqueness issue [46]. We also note that there is an additional pre-trained quantization network (approximated using a smooth function) when the IDE is connected to FSE. The quantization network transfers input continuous values to 8 possible discrete values of reflection phases for realistic implementation of the programmable metasurfaces with FPGA. More details of the DNN architecture can be found in Sec. 1 in the Supplementary Materials.

## 2.2 Design of programmable metasurfaces and implementation of on-site training

To obtain the experimental training data, we design and fabricate a programmable metasurface consisting of $20 \times 20$ unit cells operating at 11GHz, in which the reflection of each column $\{\varphi_i\}$ can be independently controlled as shown in Fig. 2(a). The reflected fields depend on the assigned phase profiles $\{\varphi_i\}$ on the 20 columns of the metasurface. By varying the reflection phase profiles rapidly in time, a large set of experimental data can be collected from the 3 probes within a short time for the DNN training. In our case, 10000 sets of randomly selected $\{\varphi_i\}$ are chosen as input to the metasurfaces, and the experimental training data $\{I_{mj}\}$ can be collected within 10 seconds.

Figure 2(b) shows the unit structure of the metasurface with detailed geometric parameters. Three copper layers are printed on two substrate layers (Rogers 4003C, relative permittivity $\varepsilon_r = 3.55$, loss tangent $tan\delta = 0.0027$) and a bonding layer (Rogers 4450F, $\varepsilon_r = 3.52$, $tan\delta = 0.004$). A varactor diode (MAVR-000120-14110P), as an active component whose capacitance changes with the bias voltage, is embedded between two metallic patches on the top layer. Two metallic vias are used to electrically connect to the negative "−" electrode in the middle layer and the positive "＋" electrode in the bottom layer, respectively. By applying different bias voltages to the varactor diode, the dipole resonance of the metasurface can be shifted in the frequency domain, leading to programmable reflection phase response at a fixed working frequency. The measured reflection phase spectrum of the metasurface at 8 different bias voltages is shown in Fig. 2(c). At the operating frequency of 11GHz indicated by the vertical orange line in the figure, we use 8 discrete phase states with 45-degree gradient covering a 315-degree range (to be set by FPGA). These phase states are assigned to the reflection phases $\{\varphi_i\}$ of the 20 independent columns to create different phase profiles. The reflection amplitudes for these 8 states have some variation (within 1.7 dB) but this limitation of the implementation has already been considered in the DNN as the network is trained directly from experimental data.

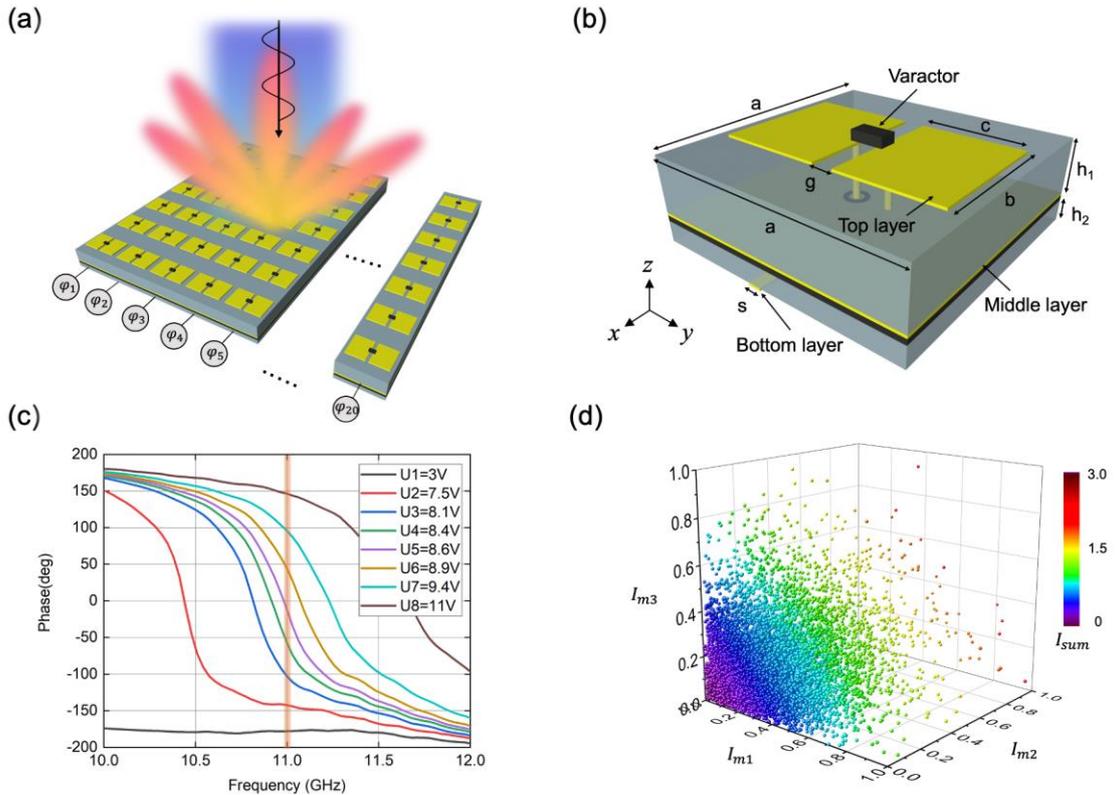

Fig. 2 Metasurface design and experimental training data collection. (a) Schematic of the programmable metasurface design. The reflection phases $\{\varphi_i\}$ for 20 columns on the metasurface can be dynamically controlled in time. Scattered fields caused by large sets of phase profiles are experimentally collected for the DNN training. (b) Unit cell structure with geometric parameters. The embedded varactor diode works as an active component, whose capacitance changes with the bias voltage, leading to the frequency shift of the dipole resonance of the metasurface. (c) Measured reflection phase of the metasurface at different bias voltages. The vertical orange line indicates the operating frequency at 11GHz. Eight discrete phase states at 11 GHz are used to control the metasurface. (d) Normalized intensities $I_{m1}$, $I_{m2}$, and $I_{m3}$ experimentally measured from the 3 probes for the DNN training process in the scenario without obstacle. 10000 sets of intensities $\{I_{mj}\}$ are collected from 10000 sets of the random reflection

phases $\{\varphi_i\}$. The color of points denotes the sum of the intensities from the 3 probes.

There is a need to investigate the possible range of the measured intensities $\{I_{mj}\}$ from the metasurface. In the scenario without the obstacle placed in front of the metasurface, we randomly generate 10000 sets of phase profiles on the metasurface and use the 3 fixed probes to experimentally measure the corresponding intensities $\{I_{mj}\}$. The intensities are plotted as three-dimensional points in Fig. 2(d). The details of experimental data collection can be found in Fig. S5 in the Supplementary Materials. We note that the $\{I_{mj}\}$ plotted in the figure are normalized by $I_{mj}/I_{max}$, where the $I_{max}$ denotes the maximum intensity received from the 3 probes in the given 10000 sets of measurements. The color of the points denotes the sum of the intensities from the 3 probes. The contour surfaces show up approximately as planes and more data points are located near the coordinate origin. The normalized intensities $I_{m1}$, $I_{m2}$, and $I_{m3}$ less than 0.6 account for 97.9%, 94.1%, and 98.8% of the total data, respectively. In the following, these data are used to train the DNN, and any target normalized intensity values are assumed to range from 0 to 0.6.

## 3. Experimental results
### 3.1 DNN training and testing without obstacle

The proposed experiment-based deep learning approach for power allocation is first demonstrated in the scenario without the obstacle. The randomly generated phase profiles $\{\varphi_i\}$ and the corresponding measured intensities $\{I_{mj}\}$ in Fig. 2(d) comprise 10000 sets of data, with 8100 sets used for training, 900 sets for validation, and the remaining 1000 sets for testing. The MSE loss function between $\{I'_j\}$ and $\{I_{mj}\}$ is used to train the FSE for 10000 epochs with Adam optimizer, and a learning rate of 0.0001, achieving the final validation loss of 0.002. The testing results of the FSE are shown in Fig. S3 in the Supplementary Materials, indicating a relatively low MSE of 0.0011 on average. Then we train the integrated DNN comprising the IDE and pre-trained FSE. 48000 configurations of the target intensities $\{I_j\}$ are randomly generated with target intensity at each probe $j$ chosen from a uniform distribution $U[0, 0.6]$, which is a reasonable range of the DNN to achieve as illustrated before from Fig. 2(d). 40500 sets of the $\{I_j\}$ are used as training data, 4500 sets as validation data, and the remaining 3000 sets as testing data. The integrated DNN is trained for 6000 epochs, using the MSE loss function between $\{I_j\}$ and $\{I'_j\}$ and an Adam optimizer with a learning rate of 0.0005, achieving the final validation loss of 0.0003 for convergence. Notably, only the weights of the IDE are optimized in the latter training process.

To test the performance of the trained DNN, we first demonstrate three special cases called "001", "101", and "000". The "001" case denotes that the metasurface can manipulate the scattered fields toward one particular probe with a strong signal while the other two probes obtain weak signals. Similarly, the "101" case shows that two probes receive strong signals while the central probe receives a weak signal. The "000" case means minimum or zero target power level for the signals to be received for all the 3 probes. As shown in the black bars in Fig. 3(a)-(c), we show the target normalized intensities $\{I_1, I_2, I_3\}$ as {0, 0, 0.55}, {0.55, 0, 0.55}, and {0, 0, 0} to the trained DNN, corresponding to the three special cases. The IDE is then used to output the reflection phases $\{\varphi_i\}$ (after quantization network) for the metasurface to fulfill the demand targets. Then the reflection phases are regarded as the input of the FSE, generating the predicted intensities $\{I'_1, I'_2, I'_3\}$ (orange bars) that agree well with the target values. To experimentally validate the network predictions, we implement the obtained reflection phases (from the IDE) to the metasurface and experimentally measure the intensities $\{I_{m1}, I_{m2}, I_{m3}\}$ from the 3 probes. As can be seen in the figure, the experimentally measured results (blue bars) match well with the targets and network predictions, showing our DNN-assisted metasurface can manipulate the scattered fields on demands to realize the three special cases, in an actual experimental setting. Particularly, these results enable the application of the programmable metasurface to deliver and damp signal receiving at different locations, pointing to applications for the metasurfaces as RISs, e.g. for a room decorated with such metasurfaces to selectively deliver signals at different locations [16]. We note that the experimental conditions remain the same for the whole training and test process.

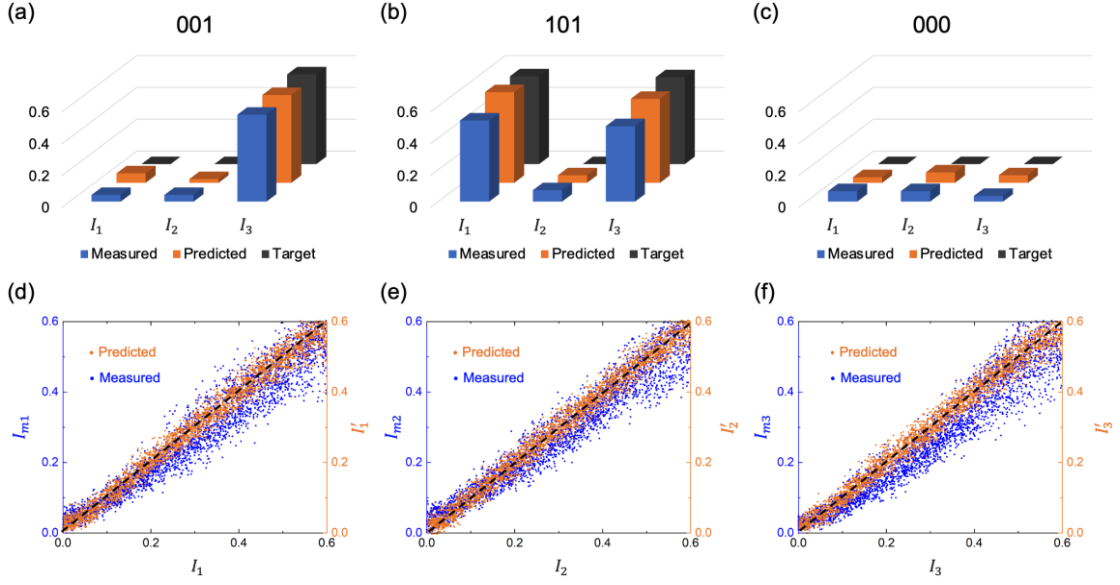

Fig. 3 Performance of DNN-assisted power allocation without obstacle. (a)-(c) Three special cases of "001", "101", and "000" for the 3 probes. The black bars, orange bars, and blue bars represent the target intensities $\{I_j\}$, the predicted intensities $\{I'_j\}$ from the DNN, and the measured intensities $\{I_{mj}\}$ from the experiment, respectively. (d)-(f) General cases for 3 probes with 3000 sets of test data. The predicted intensities $\{I'_j\}$ (orange points) and the measured intensities $\{I_{mj}\}$ (blue points) are both plotted against the target intensities $\{I_j\}$ (horizontal axis). The black dashed line is plotted for reference. The closer the data point is to the reference line, the smaller the error from the target value.

To evaluate the overall performance, our system can arbitrarily control the allocated power to target values within the reasonable range as shown in Fig. 3(d)-(f). We input the remaining 3000 testing sets of $\{I_1, I_2, I_3\}$ as the target intensities to the trained DNN and obtained the 3000 sets of $\{\varphi_i\}$ and predicted intensities $\{I'_1, I'_2, I'_3\}$. In Fig. 3(d)-(f), the horizontal axes and right-hand vertical axes denote the target intensities $\{I_1, I_2, I_3\}$ and predicted intensities $\{I'_1, I'_2, I'_3\}$, respectively. We observe that 3000 orange data points show a linear distribution around the dashed reference lines $I'_j = I_j$, showing that the DNN has been well-trained to predict the intensities of the three probes according to the input targets. The mean squared errors (MSEs) between $\{I_1, I_2, I_3\}$ and $\{I'_1, I'_2, I'_3\}$ are calculated and found to be $0.61 \times 10^{-3}$, $0.52 \times 10^{-3}$, $0.59 \times 10^{-3}$ for the 3 probes. Next, we evaluate the performance in an actual experimental test. The 3000 sets of $\{\varphi_i\}$ are implemented by the metasurface, and the corresponding measured intensities $\{I_{m1}, I_{m2}, I_{m3}\}$ are plotted against the target intensities. As expected, the measured results denoted by blue points are distributed linearly around the dashed reference lines $I_{mj} = I_j$, indicating the system can control the allocated power at the 3 probes to target values. The MSEs for measured results are obtained as $2.4 \times 10^{-3}$, $2.1 \times 10^{-3}$, $3.1 \times 10^{-3}$ for the 3 probes, respectively. The errors for predicted and measured results may come from limited training samples, phase quantization errors, and noisy data acquisition.

### 3.2 On-site updated DNN with obstacle

The above results have demonstrated that our DNN-assisted metasurface is capable of controlling the power to specific receivers on demand in the scenario without obstacles. Normally, a well-trained DNN for the specific scenario may fail to work after the ambient conditions change (the emergence of obstacles, for example). However, our experiment-based DNN can simply adapt to the changed ambient conditions because the DNN can be retrained using experimental data that can be collected within a short time and updated periodically. To demonstrate the adaptivity of the system, a metal frame obstacle is added in between the metasurface and the three probes as shown in Fig. 1. We input the same 3000 testing sets of $\{I_1, I_2, I_3\}$ to the previous DNN (trained without obstacle) and obtain the $\{\varphi_i\}$. By implementing the 3000 sets of $\{\varphi_i\}$ on the metasurface, we measure the corresponding intensities $\{I_{m1}, I_{m2}, I_{m3}\}$ and plot them with the target $\{I_1, I_2, I_3\}$ as shown in Fig. 4(a)-(c). For (a) and (b), the measured data points deviate below the dashed reference lines $I_{mj} = I_j$, which means the signals transmitted to these two probes are blocked or scattered away by the added obstacle. For (c), the measured results show a poor linear correlation with target values affected by the appearance of the obstacle. The MSEs are $7.5 \times 10^{-3}$, $22 \times 10^{-3}$, $4.6 \times 10^{-3}$ for the Fig. 4(a)-(c) respectively, showing larger errors compared with the case without obstacle in Fig. 3(d)-(f). Therefore, the original DNN trained without the obstacle performs poorly under the changed ambient conditions.

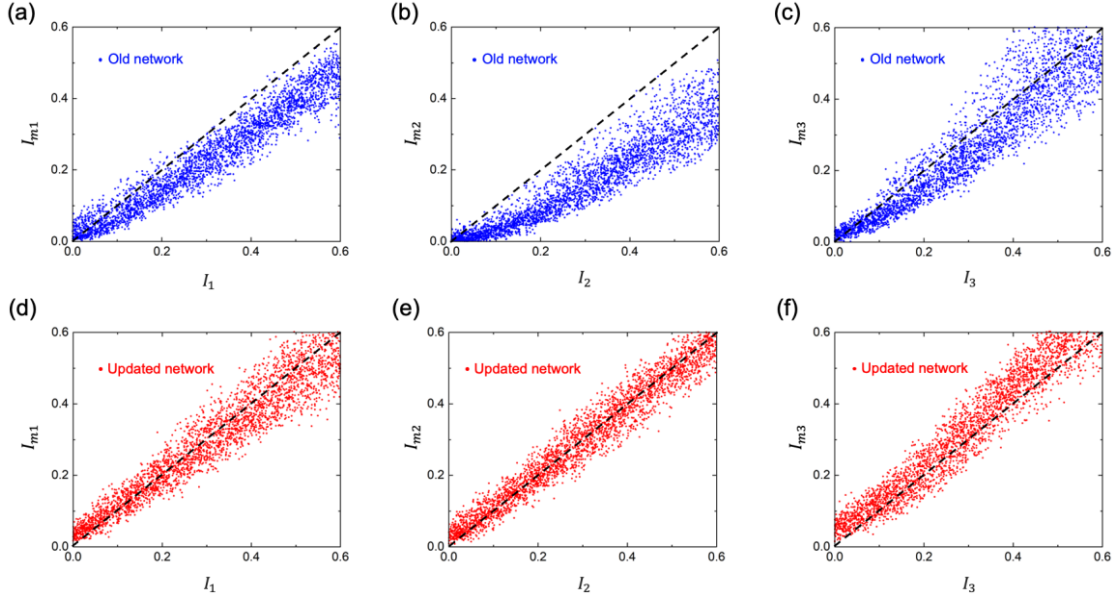

Fig. 4 Performance of DNN-assisted adaptive power allocation with an obstacle. (a)-(c) The measured intensities $\{I_{mj}\}$ against target intensities $\{I_j\}$ for the 3 probes, using the previous DNN trained without an obstacle. The data points deviate from the dashed reference line, showing the previous DNN fails to work after adding an obstacle. (d)-(f) The measured intensities $\{I_{mj}\}$ with target intensities $\{I_j\}$ using the on-site updated DNN trained with the obstacle. The data points return to the reference dashed line, showing the experiment-based DNN is adapted to the changed ambient conditions.

To adapt to the new ambient condition, we collect again experimental data and redo the training process as illustrated earlier with around 10 mins (10 seconds for experimental data collection and 9 mins for DNN training). Then, we update the DNN to suit the new scenario without knowing the material properties or structure parameters of the added obstacle. With the updated DNN, we input the same 3000 sets of testing $\{I_1, I_2, I_3\}$ and measure the intensities $\{I_{m1}, I_{m2}, I_{m3}\}$. As shown in Fig. 4(d)-(f), the measured data points restore a linear distribution around the dashed reference lines with lower MSEs of $2.6 \times 10^{-3}$, $1.4 \times 10^{-3}$, $3.8 \times 10^{-3}$. It is obvious that the updated DNN adapts to a changed environment and works well in the complex scenario with an obstacle, showing that our proposed experiment-based DNN can achieve power allocation in complex environments, and the on-site rapid DNN retraining capability makes our system adaptable to changed environments.

## 4. Discussion and Conclusion

In this work, we use 3 probes with specific locations to collect the experimental training data for the DNN construction. Our scheme also allows for the control of scattered fields in other locations by adding more probes, depending on the number of target users. Furthermore, at the current stage, our system manipulates field patterns in the horizontal plane, as our metasurface only has the degree of freedom to control the phase profiles along the y direction (each column is independently controlled). To further enhance the system's capabilities, power allocation with higher degrees of freedom in space can be achieved by independently controlling each unit cell of the metasurface in two transverse dimensions.

In summary, we have proposed the experiment-based DNN approach for power allocation enabled by a programmable metasurface. We directly train the DNN using experimental data, circumventing the need for complex modeling and computationally intensive simulations as training data. The experimental data can inherently incorporate complex factors that can be challenging to simulate or model, leading to more reliable and robust DNN results. Our experimental results demonstrate that experiment-based DNN can effectively control power transmitted towards multiple receivers and can adapt to the changed environments through its on-site data-collecting and fast-retraining capability. Our work provides valuable insights into the potential of leveraging real-world data for more accurate and efficient metasurface designs for intelligent and energy-efficient wireless communication in complex indoor environments.

## Acknowledgment

This work is supported by the Hong Kong Research Grants Council with project number R6015-18 and C6012-20G.

## Reference

1. Shelby, R.A., D.R. Smith, and S. Schultz, *Experimental verification of a negative index of refraction.* science, 2001. **292**(5514): p. 77-79.
2. Valentine, J., et al., *Three-dimensional optical metamaterial with a negative refractive index.* nature, 2008. **455**(7211): p. 376-379.
3. Landy, N.I., et al., *Perfect metamaterial absorber.* Physical review letters, 2008. **100**(20): p. 207402.
4. Fang, N., et al., *Sub-diffraction-limited optical imaging with a silver superlens.* science, 2005.

**308**(5721): p. 534-537.
5. Schurig, D., et al., *Metamaterial electromagnetic cloak at microwave frequencies.* Science, 2006. **314**(5801): p. 977-980.
6. Ergin, T., et al., *Three-dimensional invisibility cloak at optical wavelengths.* science, 2010. **328**(5976): p. 337-339.
7. Chen, H.-T., et al., *Active terahertz metamaterial devices.* Nature, 2006. **444**(7119): p. 597-600.
8. Fan, K. and W.J. Padilla, *Dynamic electromagnetic metamaterials.* Materials Today, 2015. **18**(1): p. 39-50.
9. Chen, K., et al., *A reconfigurable active Huygens' metalens.* Advanced materials, 2017. **29**(17): p. 1606422.
10. He, Q., S. Sun, and L. Zhou, *Tunable/reconfigurable metasurfaces: physics and applications.* Research, 2019. **2019**.
11. Shaltout, A.M., V.M. Shalaev, and M.L. Brongersma, *Spatiotemporal light control with active metasurfaces.* Science, 2019. **364**(6441): p. eaat3100.
12. Zhang, Y., et al., *Electrically reconfigurable non-volatile metasurface using low-loss optical phase-change material.* Nature Nanotechnology, 2021. **16**(6): p. 661-666.
13. Gu, T., et al., *Reconfigurable metasurfaces towards commercial success.* Nature Photonics, 2023. **17**(1): p. 48-58.
14. Cui, T.J., et al., *Coding metamaterials, digital metamaterials and programmable metamaterials.* Light: science & applications, 2014. **3**(10): p. e218-e218.
15. Huang, C., et al., *Reconfigurable intelligent surfaces for energy efficiency in wireless communication.* IEEE transactions on wireless communications, 2019. **18**(8): p. 4157-4170.
16. Liu, Y., et al., *Reconfigurable intelligent surfaces: Principles and opportunities.* IEEE communications surveys & tutorials, 2021. **23**(3): p. 1546-1577.
17. Saifullah, Y., et al., *Recent progress in reconfigurable and intelligent metasurfaces: A comprehensive review of tuning mechanisms, hardware designs, and applications.* Advanced Science, 2022. **9**(33): p. 2203747.
18. Wang, S.R., et al., *Asynchronous Space‐Time‐Coding Digital Metasurface.* Advanced Science, 2022. **9**(24): p. 2200106.
19. Wu, G.-B., et al., *Sideband-free space–time-coding metasurface antennas.* Nature Electronics, 2022: p. 1-12.
20. Zhang, L., et al., *Space-time-coding digital metasurfaces.* Nature communications, 2018. **9**(1): p. 1-11.
21. Dai, J.Y., et al., *High‐efficiency synthesizer for spatial waves based on space‐time‐coding digital metasurface.* Laser & Photonics Reviews, 2020. **14**(6): p. 1900133.
22. Zhang, L., et al., *A wireless communication scheme based on space-and frequency-division multiplexing using digital metasurfaces.* Nature electronics, 2021. **4**(3): p. 218-227.
23. Zhang, L., et al., *Breaking reciprocity with space‐time‐coding digital metasurfaces.* Advanced materials, 2019. **31**(41): p. 1904069.
24. Li, L., et al., *Electromagnetic reprogrammable coding-metasurface holograms.* Nature communications, 2017. **8**(1): p. 197.
25. Venkatesh, S., et al., *A high-speed programmable and scalable terahertz holographic metasurface based on tiled CMOS chips.* Nature electronics, 2020. **3**(12): p. 785-793.
26. Qu, G., et al., *Reprogrammable meta-hologram for optical encryption.* Nature communications,

2020. **11**(1): p. 5484.

27. Zhang, J., et al., *Generation of time-varying orbital angular momentum beams with space-time-coding digital metasurface.* Advanced Photonics, 2023. **5**(3): p. 036001.

28. Bai, X., et al., *High-efficiency transmissive programmable metasurface for multimode OAM generation.* Advanced Optical Materials, 2020. **8**(17): p. 2000570.

29. Jensen, J.S. and O. Sigmund, *Topology optimization for nano-photonics.* Laser & Photonics Reviews, 2011. **5**(2): p. 308-321.

30. Piggott, A.Y., et al., *Inverse design and demonstration of a compact and broadband on-chip wavelength demultiplexer.* Nature Photonics, 2015. **9**(6): p. 374-377.

31. Yang, H., et al., *A programmable metasurface with dynamic polarization, scattering and focusing control.* Scientific reports, 2016. **6**(1): p. 1-11.

32. Su, J., et al., *Ultrawideband radar cross-section reduction by a metasurface based on defect lattices and multiwave destructive interference.* Physical Review Applied, 2019. **11**(4): p. 044088.

33. Liu, Z., et al., *Generative model for the inverse design of metasurfaces.* Nano letters, 2018. **18**(10): p. 6570-6576.

34. Qiu, T., et al., *Deep learning: a rapid and efficient route to automatic metasurface design.* Advanced Science, 2019. **6**(12): p. 1900128.

35. Jiang, J., M. Chen, and J.A. Fan, *Deep neural networks for the evaluation and design of photonic devices.* Nature Reviews Materials, 2021. **6**(8): p. 679-700.

36. Jiang, J., et al., *Free-form diffractive metagrating design based on generative adversarial networks.* ACS nano, 2019. **13**(8): p. 8872-8878.

37. Luo, Y., et al., *Design of task-specific optical systems using broadband diffractive neural networks.* Light: Science & Applications, 2019. **8**(1): p. 112.

38. Ma, W., et al., *Deep learning for the design of photonic structures.* Nature Photonics, 2021. **15**(2): p. 77-90.

39. Ding, H., et al., *Deep learning enables accurate sound redistribution via nonlocal metasurfaces.* Physical Review Applied, 2021. **16**(6): p. 064035.

40. Fan, Z., et al., *Homeostatic neuro-metasurfaces for dynamic wireless channel management.* Science Advances, 2022. **8**(27): p. eabn7905.

41. Lin, H., et al., *Machine-learning-assisted inverse design of scattering enhanced metasurface.* Optics Express, 2022. **30**(2): p. 3076-3088.

42. Jiang, Y., et al., *Programmable metasurface RCS prediction under obstacles based on DNN.* Frontiers in Materials, 2022. **9**: p. 996956.

43. Noh, J., et al., *Reconfigurable reflective metasurface reinforced by optimizing mutual coupling based on a deep neural network.* Photonics and Nanostructures-Fundamentals and Applications, 2022. **52**: p. 101071.

44. Guo, Z., et al., *Physics-assisted generative adversarial network for X-ray tomography.* Optics Express, 2022. **30**(13): p. 23238-23259.

45. Liu, R., et al., *Recovery of continuous 3d refractive index maps from discrete intensity-only measurements using neural fields.* Nature Machine Intelligence, 2022. **4**(9): p. 781-791.

46. Liu, D., et al., *Training deep neural networks for the inverse design of nanophotonic structures.* Acs Photonics, 2018. **5**(4): p. 1365-1369.